%% file: main.tex
\documentclass[a4paper,11pt]{article}
\usepackage{jheppub} 
\usepackage{lineno}

\usepackage[utf8]{inputenc}
\usepackage{enumitem}
\usepackage{amsmath}
\usepackage{xspace}
\usepackage{subcaption}
\usepackage{comment}
\usepackage{wrapfig}
\usepackage{cprotect}

\title{Searching for Quirks at LHCb}

\author{Xabier Cid Vidal,$^a$}
\author{Miguel Fernández Gómez,$^a$}
\author{Matthew Low,}
\author{Alejandro Novo Cal,$^a$}
\author{Yuhsin Tsai,$^b$}
\author{Carlos Vázquez Sierra$^c$}

\affiliation[a]{IGFAE, Universidade de Santiago de Compostela, Santiago de Compostela, Spain}
\affiliation[b]{Department of Physics and Astronomy, University of Notre Dame, South Bend, USA}
\affiliation[c]{Departmento de Física y Ciencias de la Tierra, Universidade da Coruña, A Coruña, Spain}

\emailAdd{xabier.cid.vidal@cern.ch}
\emailAdd{miguel.fernandez.gomez@cern.ch}
\emailAdd{matthew.w.low@gmail.com}
\emailAdd{alejandro.novo.cal@cern.ch}
\emailAdd{ytsai3@nd.edu}
\emailAdd{carlos.vazquez@cern.ch}

\abstract{%
Quirks are heavy particles connected by a flux tube from a hidden confining force that remain weakly constrained in large regions of their parameter space.  This flux tube acts as a string that, at short enough distance, stretches as the quirk pair separates, then pulls the pair back together leading to interesting dynamics.  We propose a novel search using the LHCb Vertex Locator (VELO), whose forward geometry and software-based trigger are uniquely suited to detecting the characteristic back-to-back, planar hit patterns produced by quirk pairs with little transverse recoil. Using detailed simulations of the VELO geometry, together with simple geometric selections, we present different sensitivity projections, demonstrating that LHCb can probe parameter regions inaccessible to existing ATLAS and CMS searches and offering a powerful, complementary path toward discovering quirks.}

\begin{document}
\maketitle
\flushbottom

\input{introduction}
\input{experimental_context}

\input{lhcb_context}
\input{analysis}

\input{backgrounds}
\input{conclusions}


\acknowledgments

We thank Fred Blanc and Stefania Vecchi for providing comments to our manuscript. The work of XCV, MFG and ANC is supported by the Spanish Research State Agency under projects PID2022-139514NA-C33, PCI2023-145984-2 and EUR2025-164824; by the “María de Maeztu” grant CEX2023-001318-M, funded by MICIU/AEI/10.13039/501100011033; and by the Xunta de Galicia (CIGUS Network of Research Centres and project 2025-CP119). The work of MFG is supported by Xunta de Galicia under Programa de axudas á etapa predoutoral Grant ED481A-2022. The work of ANC is supported by the Spanish Science Ministry through the FPU program (FPU24/01899).
The work of YT is supported by the NSF Grant PHY-2412701. YT would also like to thank the
Tom and Carolyn Marquez Chair Fund for its generous support. The work of CVS is supported by the Spanish Research State Agency under projects PID2022-139514NA-C33 and RYC2023-043804-I (funded by MICIU/AEI/10.13039/501100011033), as well as by the Xunta de Galicia under projects ED431F 2025/27 and ED431B 2025/22, and by the InTalent UDC-Inditex program.

\bibliographystyle{JHEP}
\bibliography{biblio.bib}
\end{document}

%% file: introduction.tex
\section{Introduction}
\label{sec:introduction}

Despite the impressive march of LHC limits to higher and higher mass scales, there is still a wide array of possible particles that are still not excluded by the data.  While very heavy particles lie beyond the LHC's ability to produce them in meaningful quantities, there are some particles that are still viable due to the lack of dedicated searches.  Quirks are such particles that lack feasible, dedicated searches.

Quirks were proposed in 1979~\cite{Okun:1979tgr,Okun:1980mu} and explored in 2008~\cite{Kang:2008ea} and are particles that are both charged under a new confining force and under the Standard Model (SM).  These quirks have masses $m_Q$ well above $\Lambda$, the confinement scale of the new confining force, and may produce striking experimental signatures.  Additionally, there are no particles charged under the new confining force with a mass below $\Lambda$.

One set of models in which quirks naturally arise are those of neutral\footnote{``Neutral'' in this context refers to addressing the SM hierarchies without introducing coloured top partners.} naturalness models~\cite{Batell:2022pzc}. In these models, particles are charged under a new hidden confining gauge group. All known implementations of neutral naturalness rely on a discrete symmetry that connects SM fields to their colour-neutral counterparts. Although these partner fields are neutral under SM colour, the quark partners are charged under a gauge group related to SM colour via the discrete symmetry.  Neutral naturalness models often feature a hidden confinement scale on the order of a few GeV and can even lack particles charged under the confining gauge group at or below the confinement scale $\Lambda$.  In addition to different realizations of neutral naturalness~\cite{Cai:2008au,Burdman:2006tz, Burdman:2008ek, Craig:2015pha, Craig:2016kue}, quirks also appear in models of dark matter~\cite{Kribs:2009fy,Asadi:2025btr}. 

When a quirk and an anti-quirk, which we will call a quirk pair, fly apart a flux tube of the hidden confining force forms between them.  In QCD, where there are particles with masses below the confining scale, the particles flying apart would pull a light quark and anti-quark from the vacuum to form two colour neutral mesons.  In the quirk scenario, the flux tube stretches until it reaches a length $\ell \sim m_Q / \Lambda^2$ and then the energy in the flux tube pulls the quirk pair back towards each other.  In particular, for $m_Q \sim \mathcal{O}(1~\mathrm{TeV})$ and $\Lambda \sim 100{-}1000~\mathrm{eV}$, the string-like force has a significant effect, giving the two particles a typical separation of order centimeters to tens of meters. As a result, they can cross each other multiple times as they are detected by different trackers, as shown in Figure~\ref{fig:quirks}.

Electrically-charged quirks, therefore, leave very unique tracks in the detector as their dynamics are influenced both by the detector's magnetic field and by the string-like force between the quirks.  Quirks that are coloured hadronize into SM colour-neutral states that may have electric charges~\cite{Farina:2017cts}.  These quirk-hadrons have the quantum numbers of the quirk along with a light QCD quark or gluon in order to produce a colour-neutral state. 


To understand the current constraints on quirks, it is useful to examine the $m_Q$ vs.~$\Lambda$ plane and select a specific type of quirk. Most analyses focus on a coloured quirk with the same SM gauge charges as an up-type quark, assuming the quirk hadronizes into an electrically charged quirk-hadron. The production cross-section scales with the size of the representation of the new confining force. Typically, comparisons in the literature assume the confining group is SU$(N_c)$ and compute limits for $N_c = 2$, which is the approach followed here.

Searches for quirks, in the regime described above, are experimentally challenging and typically fall within the domain of Large Hadron Collider (LHC) experiments~\cite{Farina:2017cts,Knapen:2017kly,Evans:2018jmd,Curtin:2025ksm,Condren:2025czc,Forsyth:2025wks,Sha:2024hzq,Feng:2024zgp,Li:2020aoq,Li:2021tsy,Li:2019wce}. The ATLAS~\cite{ATLAS} and CMS~\cite{CMS} experiments are naturally sensitive to this type of exotic signature and, as we shall see, are already constraining parts of the quirk parameter space through monojet or Heavy Stable Charged Particles (HSCP) searches. However, in this paper we propose a novel approach, which relies on the use of the Vertex Locator (VELO) detector \cite{Akiba:2024now} of the LHCb experiment \cite{LHCb:2008vvz}.

A key advantage of LHCb is its instrumented coverage of the forward region, which complements the central acceptance of ATLAS and CMS. Standard searches in central detectors often rely on the quirk--anti-quirk system recoiling against initial-state radiation (ISR) to generate sufficient transverse momentum for triggering; however, this requirement significantly suppresses the signal rate. In the absence of hard ISR, the quirk--anti-quirk system is expected to oscillate around the beamline, primarily directed into the forward or backward regions due to the longitudinal boost of the center-of-mass frame.

Crucially, the VELO, with its exceptional spatial resolution and proximity to the interaction point, might be uniquely capable of resolving the complex topologies associated to quirks. Furthermore, the LHCb trigger system—which, as we will discuss in detail later, operates with a fully software-based architecture allowing access to full event information—permits the flexible design of dedicated selections to capture these anomalous signatures that might otherwise be discarded by standard hardware triggers. Figure \ref{fig:quirks} shows two examples of how the trajectory of quirks could be reconstructed at the VELO detector.

The structure of this paper is as follows. In Section \ref{sec:exp_context}, we review the current experimental status and existing constraints in the search for quirks. Section \ref{sec:lhcb_context} describes the LHCb experiment and the VELO detector, elucidating why this environment is uniquely suited for detecting quirk signatures. We then outline the proposed analysis strategy and event selection criteria in Section \ref{sec:analysis}, while backgrounds are described in Section \ref{sec:bkgs}. Finally, we present our results and conclusions in Section \ref{sec:conclusions}.

\begin{figure}
    \centering
    \includegraphics[width=0.48\linewidth]{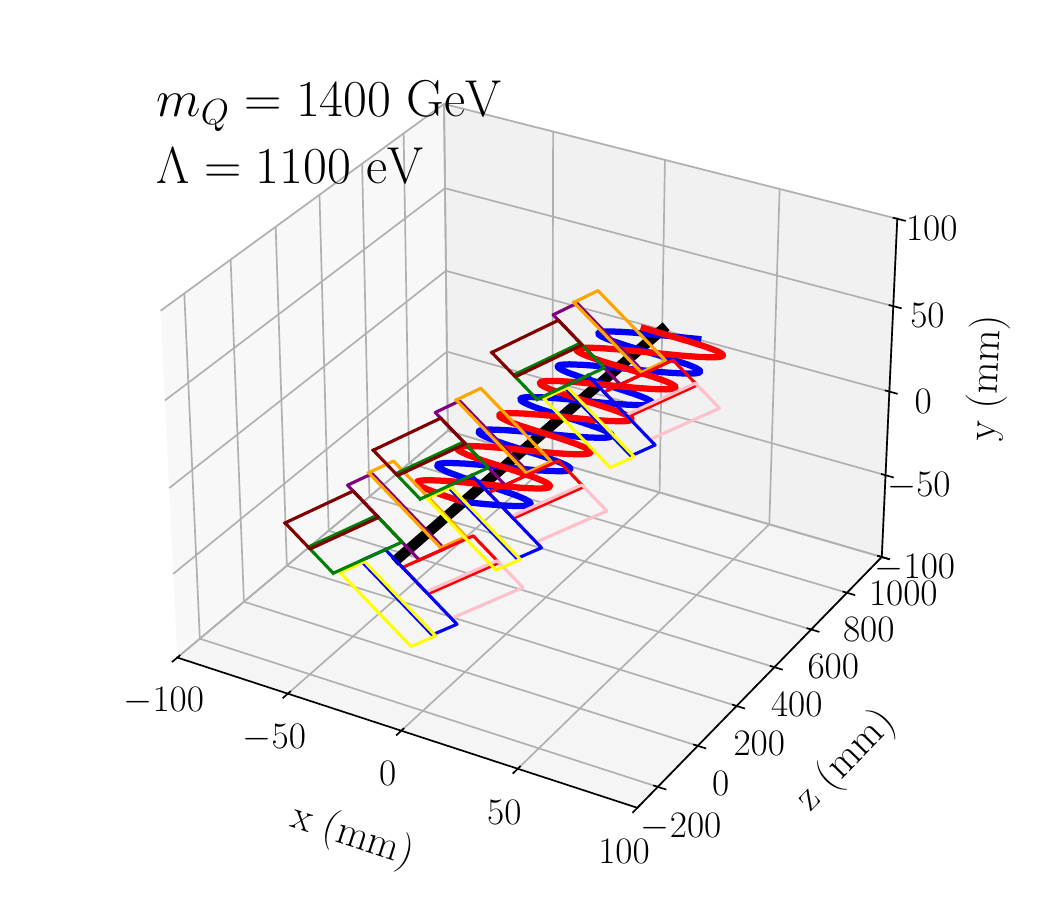}
    \includegraphics[width=0.48\linewidth]{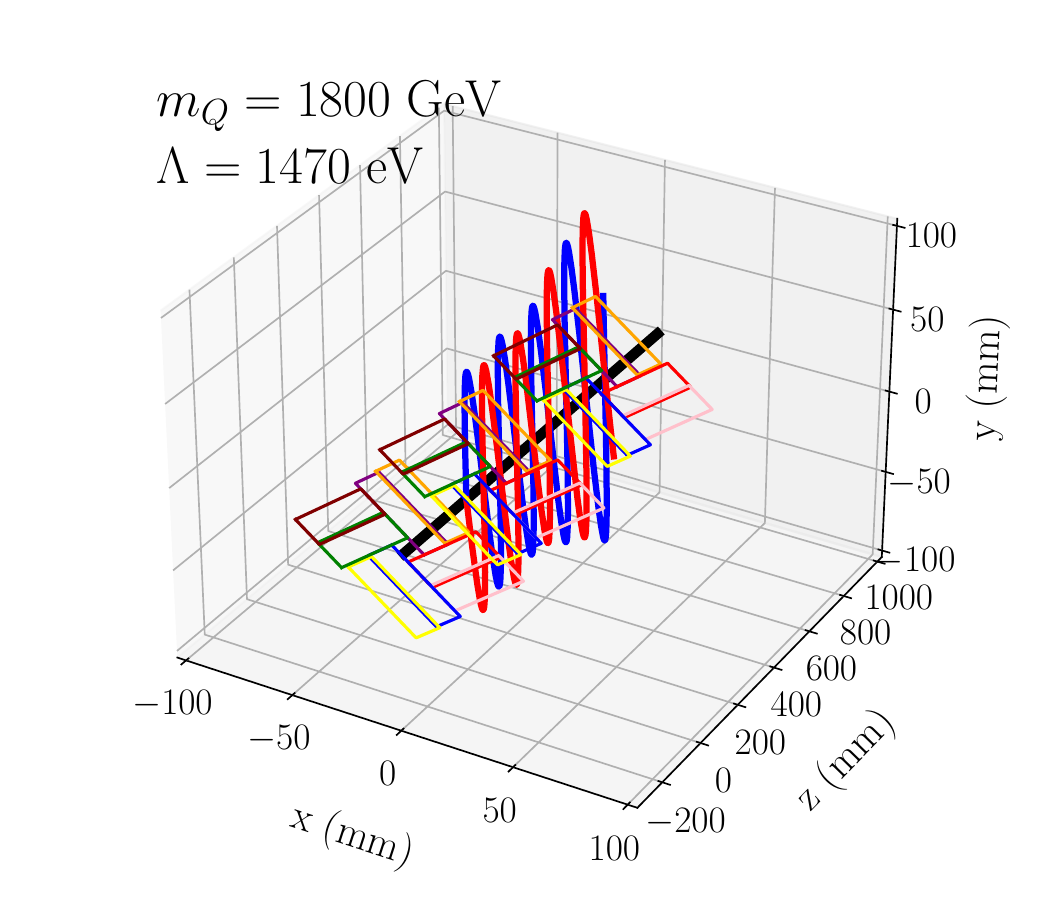}
\cprotect\caption{Example trajectories of quirk-antiquirk pairs in the VELO. Three VELO stations, centered at the origin of the $x$ and $y$ axes and shown as a series of colored rectangles parallel to one another and perpendicular to the $z$-direction, are included for reference in each case. 
\end{figure}

%% file: experimental_context.tex
\section{Existing searches}
\label{sec:exp_context}

The experimental phenomenology of quirks at collider experiments is defined by the macroscopic properties of the infracolor string connecting the pair, which, as explained, depends critically on the confinement scale $\Lambda$ relative to the quirk mass $m_Q$. Consequently, the search strategy at the LHC must vary across the parameter space, leveraging different measurements and types of searches to place bounds on these exotic states.

In Reference~\cite{Farina:2017cts}, the authors reinterpret existing LHC analyses within the quirk parameter space. When $\Lambda$ is relatively small, the string-like force between quirks has minimal influence, and their behaviour resembles that of HSCPs. Because HSCP searches are highly sensitive, they impose stringent constraints on quirks in this regime. For larger values of $\Lambda$, monojet searches, which rely on detecting missing energy and initial-state radiation, can be adapted to constrain quirks. However, these searches face challenges from significant backgrounds and reduced signal rates, resulting in weaker constraints. The reinterpretation draws on HSCP searches performed by CMS \cite{CMS:2016ybj} and monojet searches conducted by ATLAS \cite{ATLAS:2016bek}.

The model from Reference ~\cite{Farina:2017cts} has been adapted as a benchmark model in quirk studies and consists of a fermionic quirk with the quantum numbers of a right-handed top quark, namely a color triplet with hypercharge $2/3$.  This is typically written as $(3, 1)_{2/3}$.  The size of the confining group is chosen to be $N_c = 2$ since this gives the minimal cross section.  The cross section scales as $\sigma \propto N_c$ because this changes the multiplicity of the final state quirks.  The $N_c$-independent contribution to the cross section is the QCD pair production rate of such a fermion.  Changing the value of $N_c$ simply rescales $m_Q$ values relative to $\Lambda$ in the bounds.  A colored quirk forms hadrons that can be charged or neutral.  Reference ~\cite{Knapen:2017kly} estimated the fraction of charged hadrons to be $30~\%$ from Pythia~\cite{Sjostrand:2014zea}, though this number can have large uncertainties.  Similar to $N_C$ this is another overall rescaling of the rate.

Reference~\cite{Knapen:2017kly} presents an alternative method for detecting quirks in regions where HSCP searches are less effective. This strategy focuses on identifying pairs of hits in the tracker that are aligned within the same plane (similar to the picture depicted in Figure~\ref{fig:quirks}), referred to as co-planar hits. To ensure sufficient triggering efficiency, the method requires the presence of a high-$p_T$ jet from initial-state radiation. Additionally, the event selection process includes a cut-off on the quirk–anti-quirk separation distance, limiting it to $1$ cm to control the background. While this cut-off helps reduce background, it also diminishes sensitivity for intermediate values of $\Lambda$.
The quirk pair may also lose substantial energy through ionisation and eventually come to rest within the detector. During idle periods of the LHC (no active collisions), these quirks can annihilate, producing distinctive signals. Such events are constrained by searches for stopped Long-Lived Particles (LLPs) or those occurring out of time (OoT). Reference~\cite{Evans:2018jmd} proposes optimisations of the CMS analysis for OoT particles \cite{CMS:2017kku} to achieve similar or better sensitivity compared to the co-planar hits approach.
Another intriguing proposal, discussed in Reference~\cite{Feng:2024zgp}, involves using delayed or slow-moving tracks in conjunction with the FASER2 detector. Although future FASER2's capabilities may allow it to match or surpass the OoT strategy in sensitivity, this approach is not discussed further since FASER 2 is not an approved experiment yet.

\begin{figure}
    \centering
    \includegraphics[width=0.7\linewidth]{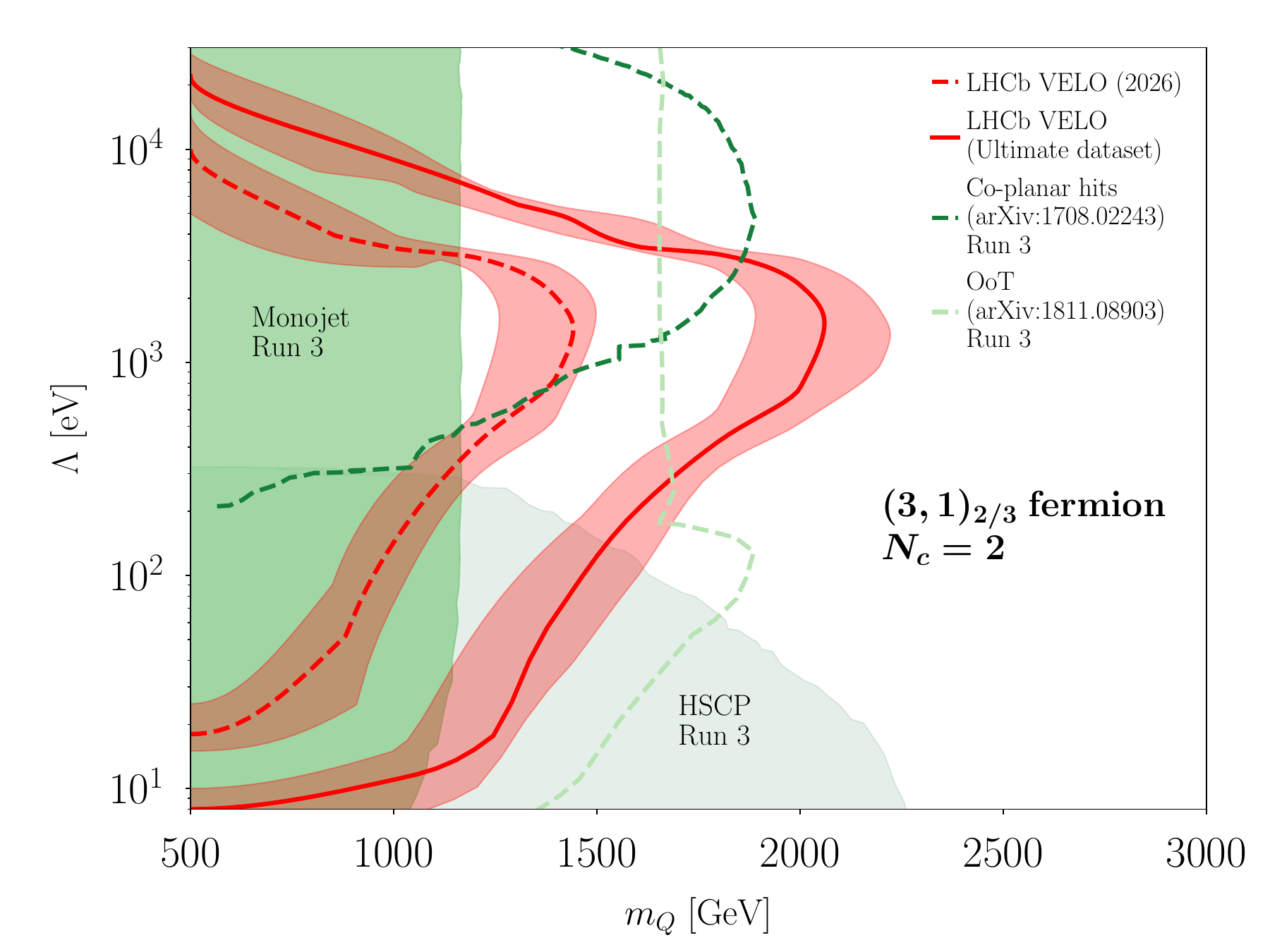}
    \caption{Projected constraints on quirks compared to expected LHCb sensitivity. The corresponding integrated luminosities are $425~\text{fb}^{-1}$, $10~\text{fb}^{-1}$ and $300~\text{fb}^{-1}$ for the ATLAS/CMS Run 3, LHCb 2026 and LHCb ultimate datasets, respectively. The red bands around the LHCb sensitivities account for a factor of 2 variation (higher or lower) in the signal production cross section, providing an indication of how the sensitivity would change if the model (or efficiencies) differed signficantly. Bounds are showed at 95\% CL. The LHCb sensitivity is obtained with a toy simulation of the VELO detector reproducing LHCb running conditions.}
    \label{fig:existing}
\end{figure}

The most relevant projected bounds discussed in this section are shown in Figure~\ref{fig:existing}, assuming a total integrated luminosity of $425~\text{fb}^{-1}$. This corresponds to the full dataset expected from ATLAS and CMS by the end of Run 3.\footnote{This comprises the $300~\text{fb}^{-1}$ already collected by each experiment, plus an additional $125~\text{fb}^{-1}$, assuming that in 2026 both experiments record the same integrated luminosity as in 2025~\cite{CMSLumiPublic,AtlasLumiRun3}.}
From the figure, we observe that despite existing limits (HSCPs and monojets) and proposed searches (co-planar hits, OoT), there remain weak constraints in the mid-$\Lambda$ range. In this regime, the LHCb experiment has the potential to set competitive bounds due to its unique detector capabilities.

%% file: lhcb_context.tex
\section{The LHCb experiment}
\label{sec:lhcb_context}

The LHCb experiment can be used to provide a unique, complementary approach in the search for quirks. LHCb has transitioned into a general-purpose detector in the forward region, having conducted several searches for LLPs \cite{LHCb:2021dyu,LHCb:2020akw,LHCb:2020ysn,LHCb:2019vmc,LHCb:2017trq,LHCb:2017xxn,LHCb:2016inz,LHCb:2016buh,LHCb:2016awg,LHCb:2014osd,LHCb:2014jgs}.
It is also essential to note that LHCb is expected to continue data-taking in the high luminosity LHC (HL-LHC) phase (Upgrade II). The ultimate integrated luminosity target for this phase is approximately $300~\text{fb}^{-1}$ \cite{LHCb:2018coe}. For the immediate future, specifically for 2026, we assume an integrated luminosity of roughly $10~\text{fb}^{-1}$, comparable to the successful $11.8~\text{fb}^{-1}$ of proton-proton collision data actually collected during the 2025 run \cite{LHCb-Outreach:2025run}.

From 2022 onwards, the LHCb trigger operates as a High Level Trigger (HLT), fully software-based and divided into two levels: HLT1 and HLT2. The HLT1, performing a rapid partial reconstruction step, is GPU-based \cite{Aaij:2019zbu}. Events passing through HLT1 are buffered on disk for subsequent filtering and storage at HLT2. At HLT2, given the buffered events, the reconstruction attains the same quality as offline. Importantly for LLP reconstruction, with this new trigger system full event information is available for decision-making in every event. Moreover, a flexible reconstruction step takes place, which can be adapted to search for non-standard LLP signatures, such as quirks.
While the strict latency constraints of the online trigger ($30~\text{MHz}$ throughput) limit the complexity of Machine Learning (ML) algorithms that can be deployed during the initial real-time selection, the system's architecture offers a critical workaround. Because the trigger buffers selected events to disk, heavy and sophisticated ML algorithms, which would be too slow for the live data stream, can be trained for and applied to these stored events during offline analysis. This allows for the identification of complex quirk signatures that would otherwise be missed by standard low-latency triggers.

The search for quirks at LHCb that we propose utilises the VELO \cite{Akiba:2024now}, a high-precision tracking detector located close to the LHC interaction point, designed to reconstruct primary and secondary vertices with exceptional accuracy. It consists of 52 modules of pixel sensors, aligned along the LHC beam pipe, spanning approximately $-300~\text{mm}$ to $750~\text{mm}$. To ensure complete angular coverage and hermeticity, the stations on the opposite sides of the beam are slightly shifted relative to each other along the beam axis ($z$-direction). This staggering allows the two halves of the detector to overlap when moved into the closed data-taking position without mechanical interference, effectively eliminating any gaps in acceptance. The VELO provides precise three-dimensional hit information, with a single-hit spatial resolution of $\mathcal{O}(10~\mu\text{m})$ \cite{Buchanan:2022iib}. A key feature of the VELO is its lack of a magnetic field, which significantly simplifies quirk trajectories.

%% file: analysis.tex
\section{Proposed LHCb analysis}
\label{sec:analysis}

\begin{figure}[t]
    \centering
    \includegraphics[width=0.49\textwidth]{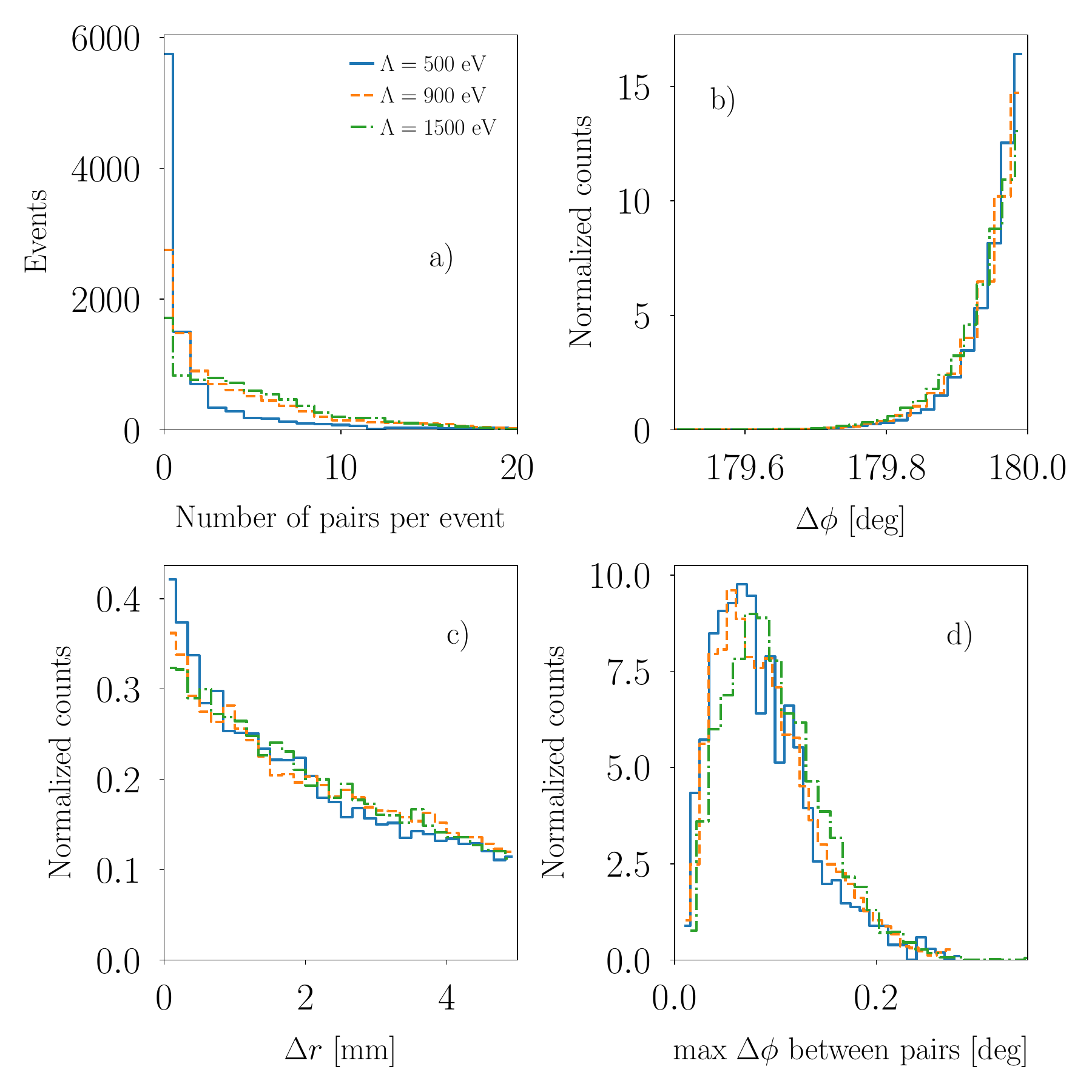}
    \includegraphics[width=0.49\textwidth]{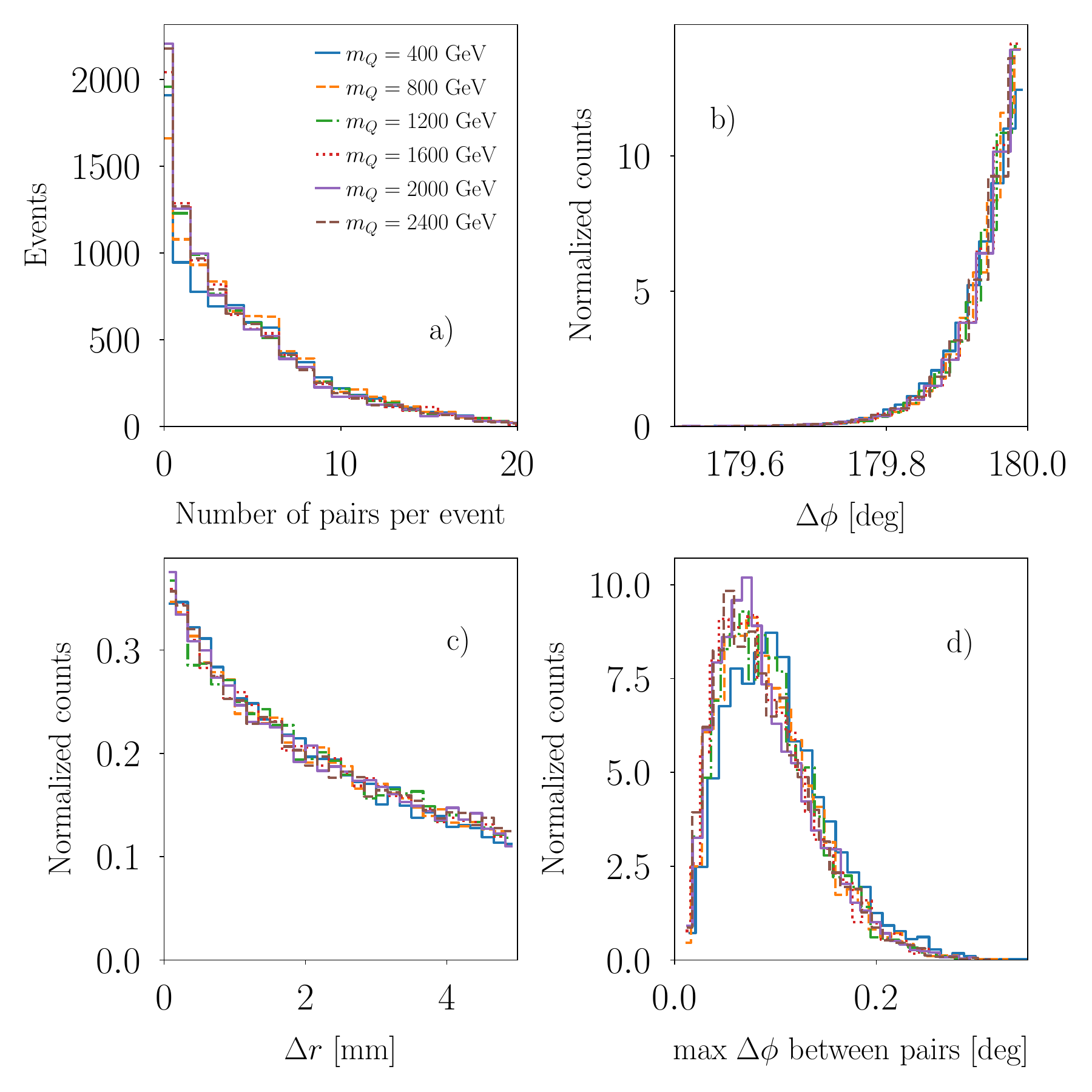}
    \caption{Characteristic kinematic features of quirk hits in the VELO detector, simulated with $m_Q=1200$ GeV and varying $\Lambda$ (left) and $\Lambda = 1100$ eV and varying $m_Q$ (right).  An electric charge of $Q_E -1$ is assumed. The panels show: a) the number of reconstructed hit pairs per event; b) the azimuthal difference $\Delta\phi$ between paired hits; c) the radial difference $\Delta r$ between paired hits; and d) the maximum $\Delta\phi$ variation across different modules. Note that the geometric correlations (b, c, d) remain mostly stable across the parameter space. The distributions are obtained with a toy simulation of the VELO detector reproducing LHCb running conditions.}
    \label{fig:all_draws}
\end{figure}

\begin{figure}[t]
    \centering
    \includegraphics[width=0.95\linewidth]{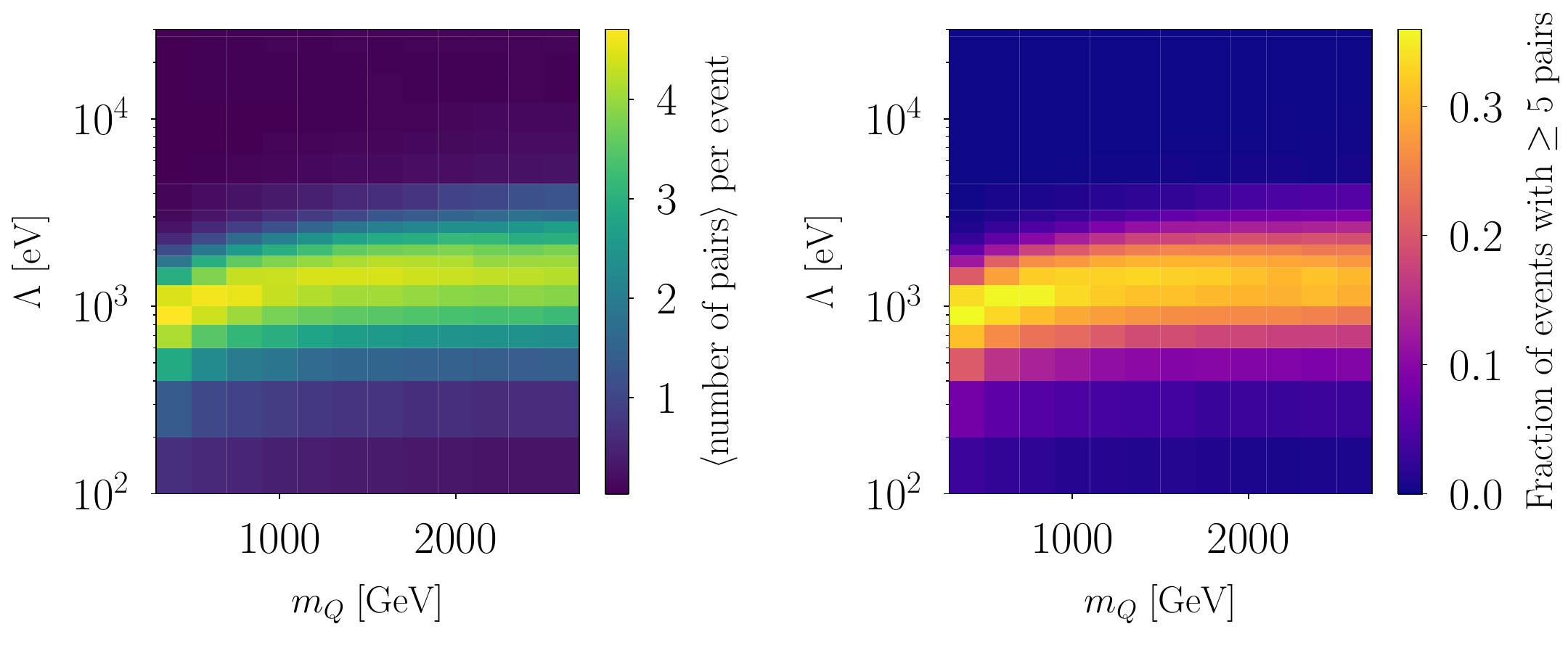}
    \caption{Projected VELO detection efficiency across the quirk parameter space ($m_Q$ vs. $\Lambda$). Left: Average number of reconstructed hit pairs per event. Right: Fraction of events satisfying the selection criterion of at least 5 reconstructed hit pairs. The efficiencies are obtained with a toy simulation of the VELO detector reproducing LHCb running conditions.}
    \label{fig:maphits}
\end{figure}

We propose that LHCb searches for quirk--anti-quirk pairs moving predominantly along the beam axis with minimal transverse recoil. To evaluate the feasibility of this search, we simulate quirk trajectories using a standalone code that generates hits according to the precise VELO geometry \cite{Coco:2683878}, explicitly accounting for the single-hit spatial resolution of approximately $10~\mu\text{m}$ \cite{Buchanan:2022iib}.
Our study uses toy simulations based on reference points~\cite{Farina:2017cts,Knapen:2017kly}, with a SM-coloured quirk (see above). This choice was made to facilitate direct comparison with previous studies. 
The signal manifests as pairs of hits in consecutive $z$-layers of the VELO detector. Unlike conventional particle tracks, which originate from the primary vertex and propagate outwards, quirk hits may not necessarily appear at monotonically increasing radii in the forward layers. However, the signal topology is distinct: the quirk and anti-quirk are expected to leave hits in modules positioned on opposite sides of the beam pipe. To reconstruct these, we require pairs of hits located in modules facing each other (see again Figure \ref{fig:quirks}). Although the VELO modules are mechanically staggered (slightly offset in the $z$-direction), the radial difference $\Delta r$ between the paired hits is expected to be small; for our selection, we require this difference to be under $5~\text{mm}$. Furthermore, the azimuthal difference $\Delta \phi$ between the hits is expected to be very close to $180^{\circ}$ (back-to-back in the transverse plane), and the maximum variation in $\phi$ between pairs across different modules is minimal, indicating a planar trajectory. As shown in Figure \ref{fig:all_draws}, these geometric definitions, particularly the planarity and back-to-back correlation, remain stable across the quirk parameter space.

While the geometric signatures are robust, the primary factor determining the selection efficiency is the total number of reconstructed hits per event. This quantity varies significantly depending on the quirk mass $m_Q$ and the confinement scale $\Lambda$, as illustrated in Figure \ref{fig:maphits}. Consequently, the hit multiplicity dictates the ultimate sensitivity of the search. The geometry of the VELO imposes natural constraints here: the sensors cover radial distances from $7$ to $40~\text{mm}$ from the beam pipe. For high $\Lambda$ values (tight confinement), the quirk and anti-quirk cannot be spatially resolved, while for very low $\Lambda$, the macroscopic separation may impede simultaneous reconstruction within the acceptance.

As highlighted in Section~\ref{sec:exp_context}, the upgraded LHCb trigger system offers a unique opportunity to accommodate these signatures. With full event information available at the software trigger level, flexible reconstruction strategies are feasible. We hypothesize that a trigger for quirks could be developed to collect data in 2026, based on simple, hit-level criteria, as those outlined above. This initial trigger would serve as a proof of concept, offering partial sensitivity and enabling refinements for Runs 4 and 5.

%% file: backgrounds.tex
\section{Backgrounds}
\label{sec:bkgs}

To validate the robustness of this selection strategy against standard model processes, we performed a study using high-statistics samples of inelastic $pp$ collisions and $b\bar{b}$ events generated with Pythia 8~\cite{Sjostrand:2014zea}, together with the same code that generates hits according to the VELO geometry used for signal. These samples were chosen to represent the dominant sources of combinatorial background and heavy-flavour activity in the forward region. The pivotal element of our proposed selection is the topological requirement that the characteristic pairs of back-to-back hits must appear in \textit{consecutive} VELO stations. While random combinatorial associations or pile-up artifacts may occasionally mimic a single back-to-back pair, they lack the longitudinal coherence required to form a sequence across multiple adjacent $z$-layers. Our studies indicate that this strict geometric condition is highly effective: when applied to the simulated samples, the vast majority of background events are rejected, with only a handful of candidates surviving the initial filter. 

A second potential background arises from isolated photon conversions in the detector material, where the electron and positron hits are typically co-planar. To mitigate this, control channels in the data \cite{Alexander:2018png} are used at LHCb to construct precise material maps that assist in rejecting such conversions. By vetoing track pairs that originate from known material dense regions (such as the RF foil or module supports), this background can be significantly suppressed.

To eliminate any remaining candidates, we propose the use of sophisticated offline reconstruction techniques. Recently, ML approaches have been suggested as a powerful tool for reconstructing full quirk trajectories \cite{Sha:2024hzq}. We anticipate that an offline ML-based tracking algorithm, trained on the specific output of the trigger selection, would provide the necessary discrimination power to suppress the background to mostly negligible levels. In the event of a signal discovery, these ML techniques would naturally extend to the characterisation of the particle, allowing for the estimation of the quirk mass and string tension.  Further refinements to ensure signal purity include the use of the Ring Imaging Cherenkov (RICH) detectors \cite{Adinolfi:2012qfa}. Similar to Heavy Stable Charged Particle (HSCP) searches at LHCb \cite{LHCb:2015ujr}, we can require that the candidate tracks produce no corresponding signal in the RICH, consistent with slow-moving or highly ionising particles. Consequently, for the sensitivity estimates presented in this paper, we assume a background-free hypothesis. 

Finally, to address systematic uncertainties associated with material interactions and tracking efficiencies, we propose the use of control channels in data, such as photon conversions. By reweighting the trajectories of electron-positron pairs to resemble quirk signatures, we can validate the detector response. The successful application of such data-driven techniques has proven effective in reducing uncertainties in previous LLP searches at LHCb \cite{LHCb:2021dyu, LHCb:2020akw, LHCb:2020ysn}.

%% file: conclusions.tex
\section{Results and conclusions}\label{sec:conclusions}

To estimate the physics reach of the proposed search, we define a statistical framework based on the signal efficiencies derived in Section~\ref{sec:analysis}. As explained above, we assume a hadronisation model where $30\%$ of quirks form charged particles, consistent with the benchmarks established in Reference~\cite{Knapen:2017kly}. For the interaction of these heavy charged states with the detector material, we adopt a simplified assumption of $100\%$ detection efficiency for quirks with electric charge $Q_E=\pm 1$, regardless of their mass. It should be noted that detailed material interaction studies, accounting for specific energy loss mechanisms, will be required to validate this assumption in a full experimental analysis. Furthermore, the initial fragmentation of the QCD color string is expected to produce a multiplicity of low-transverse-momentum hadrons; while our strict geometric selection is anticipated to be robust against these additional soft tracks, future studies utilizing a complete detector simulation must explicitly evaluate their potential to induce combinatorial confusion during pattern recognition and quantify any resulting impact on the signal sensitivity.

The signal yield is calculated based on the selection requirement of at least five VELO stations containing back-to-back hits at consistent $\phi$ angles. We assume that the combination of the strict geometric selection and the proposed offline ML filtering reduces the background contribution to negligible levels. Consequently, exclusion limits are set using the Feldman-Cousins method \cite{Feldman:1997qc} at $95\%$ confidence level (CL). We provide sensitivity projections for two scenarios: an integrated luminosity of $10~\text{fb}^{-1}$, corresponding to the expected dataset for the 2026 run, and an ultimate scenario of $300~\text{fb}^{-1}$ representing the full HL-LHC dataset for LHCb.

The resulting sensitivity contours in the $m_Q$--$\Lambda$ plane are summarised in Figure \ref{fig:existing}. These projections suggest that LHCb could achieve unprecedented sensitivity, covering significant portions of parameter space potentially inaccessible to other experiments, particularly in the $\Lambda \sim 1000$ eV region. The search proposed here for LHCb should remain robust for Run 4 and holds excellent prospects for Run 5 (Upgrade II), particularly given the enhanced timing capabilities of the future 4D VELO upgrade \cite{deAguiarFrancisco:2024bzg}.

In conclusion, the unique geometric configuration of the LHCb VELO, combined with a flexible software-based trigger, provides a powerful and distinct handle to explore uncharted regions of the quirk parameter space. This approach offers high complementarity to the central detector searches performed by ATLAS and CMS. We strongly encourage the development of dedicated trigger lines for these signatures during the remainder of Run 3 and for future data-taking campaigns at LHCb. Moving forward, further studies will be necessary to fully validate the background-free hypothesis, specifically through the implementation of full quirk trajectory reconstruction algorithms and the inclusion of particle identification information from the RICH subdetectors.

%% file: biblio.bib
@article{Asadi:2025btr,
    author = "Asadi, Pouya and Kribs, Graham D. and Luty, Markus A.",
    title = "{Quirks Live in Cool Universes}",
    eprint = "2512.20696",
    archivePrefix = "arXiv",
    primaryClass = "hep-ph",
    reportNumber = "CERN-TH-2025-243",
    month = "12",
    year = "2025"
}

@article{Alexander:2018png,
    author = "Alexander, M. and others",
    title = "{Mapping the material in the LHCb vertex locator using secondary hadronic interactions}",
    eprint = "1803.07466",
    archivePrefix = "arXiv",
    primaryClass = "physics.ins-det",
    reportNumber = "CERN-LHCb-DP-2018-002, CERN-LHCB-DP-2018-002",
    doi = "10.1088/1748-0221/13/06/P06008",
    journal = "JINST",
    volume = "13",
    number = "06",
    pages = "P06008",
    year = "2018"
}

@article{Sjostrand:2014zea,
    author = {Sj\"ostrand, Torbj\"orn and Ask, Stefan and Christiansen, Jesper R. and Corke, Richard and Desai, Nishita and Ilten, Philip and Mrenna, Stephen and Prestel, Stefan and Rasmussen, Christine O. and Skands, Peter Z.},
    title = "{An introduction to PYTHIA 8.2}",
    eprint = "1410.3012",
    archivePrefix = "arXiv",
    primaryClass = "hep-ph",
    reportNumber = "LU-TP-14-36, MCNET-14-22, CERN-PH-TH-2014-190, FERMILAB-PUB-14-316-CD, DESY-14-178, SLAC-PUB-16122",
    doi = "10.1016/j.cpc.2015.01.024",
    journal = "Comput. Phys. Commun.",
    volume = "191",
    pages = "159--177",
    year = "2015"
}

@article{Aaij:2019zbu,
    author = "Aaij, R. and others",
    title = "{Allen: A high level trigger on GPUs for LHCb}",
    eprint = "1912.09161",
    archivePrefix = "arXiv",
    primaryClass = "physics.ins-det",
    doi = "10.1007/s41781-020-00039-7",
    journal = "Comput. Softw. Big Sci.",
    volume = "4",
    number = "1",
    pages = "7",
    year = "2020"
}

@article{LHCb:2021dyu,
    author = {{LHCb collaboration}},
    collaboration = "$\mathrm{LHCb}$",
    title = "{Search for massive long-lived particles decaying semileptonically at $\sqrt(s)=13$ TeV}",
    eprint = "2110.07293",
    archivePrefix = "arXiv",
    primaryClass = "hep-ex",
    reportNumber = "LHCb-PAPER-2021-028, CERN-EP-2021-186",
    month = "10",
    year = "2021"
}

@article{ATLAS,
    author = "Aad, G. and others",
    collaboration = "ATLAS",
    title = "{The ATLAS Experiment at the CERN Large Hadron Collider}",
    doi = "10.1088/1748-0221/3/08/S08003",
    journal = "JINST",
    volume = "3",
    pages = "S08003",
    year = "2008"
}

@article{CMS,
    author = "Chatrchyan, S. and others",
    collaboration = "CMS",
    title = "{The CMS Experiment at the CERN LHC}",
    doi = "10.1088/1748-0221/3/08/S08004",
    journal = "JINST",
    volume = "3",
    pages = "S08004",
    year = "2008"
}

@article{LHCb:2020akw,
    author = {{LHCb collaboration}},
    collaboration = "$\mathrm{LHCb}$",
    title = "{Search for long-lived particles decaying to $e^\pm \mu^\mp \nu$}",
    eprint = "2012.02696",
    archivePrefix = "arXiv",
    primaryClass = "hep-ex",
    reportNumber = "LHCb-PAPER-2020-027, CERN-EP-2020-212",
    doi = "10.1140/epjc/s10052-021-08994-0",
    journal = "Eur. Phys. J. C",
    volume = "81",
    number = "3",
    pages = "261",
    year = "2021"
}

@article{LHCb:2020ysn,
    author = {{LHCb collaboration}},
    collaboration = "$\mathrm{LHCb}$",
    title = "{Searches for low-mass dimuon resonances}",
    eprint = "2007.03923",
    archivePrefix = "arXiv",
    primaryClass = "hep-ex",
    reportNumber = "LHCb-PAPER-2020-013, CERN-EP-2020-114",
    doi = "10.1007/JHEP10(2020)156",
    journal = "JHEP",
    volume = "10",
    pages = "156",
    year = "2020"
}

@article{LHCb:2019vmc,
    author = {{LHCb collaboration}},
    collaboration = "$\mathrm{LHCb}$",
    title = "{Search for $A'\to\mu^+\mu^-$ Decays}",
    eprint = "1910.06926",
    archivePrefix = "arXiv",
    primaryClass = "hep-ex",
    reportNumber = "LHCb-PAPER-2019-031, CERN-EP-2019-212",
    doi = "10.1103/PhysRevLett.124.041801",
    journal = "Phys. Rev. Lett.",
    volume = "124",
    number = "4",
    pages = "041801",
    year = "2020"
}

@article{LHCb:2017trq,
    author = {{LHCb collaboration}},
    collaboration = "$\mathrm{LHCb}$",
    title = "{Search for Dark Photons Produced in 13 TeV $pp$ Collisions}",
    eprint = "1710.02867",
    archivePrefix = "arXiv",
    primaryClass = "hep-ex",
    reportNumber = "LHCB-PAPER-2017-038, CERN-EP-2017-248",
    doi = "10.1103/PhysRevLett.120.061801",
    journal = "Phys. Rev. Lett.",
    volume = "120",
    number = "6",
    pages = "061801",
    year = "2018",
    addendum = "(800 authors)"
}

@article{LHCb:2017xxn,
    author = {{LHCb collaboration}},
    collaboration = "$\mathrm{LHCb}$",
    title = "{Updated search for long-lived particles decaying to jet pairs}",
    eprint = "1705.07332",
    archivePrefix = "arXiv",
    primaryClass = "hep-ex",
    reportNumber = "LHCB-PAPER-2016-065, CERN-EP-2017-083",
    doi = "10.1140/epjc/s10052-017-5178-x",
    journal = "Eur. Phys. J. C",
    volume = "77",
    number = "12",
    pages = "812",
    year = "2017"
}

@article{LHCb:2016awg,
    author = {{LHCb collaboration}},
    collaboration = "$\mathrm{LHCb}$",
    title = "{Search for long-lived scalar particles in $B^+ \to K^+ \chi (\mu^+\mu^-)$ decays}",
    eprint = "1612.07818",
    archivePrefix = "arXiv",
    primaryClass = "hep-ex",
    reportNumber = "CERN-EP-2016-302, LHCB-PAPER-2016-052",
    doi = "10.1103/PhysRevD.95.071101",
    journal = "Phys. Rev. D",
    volume = "95",
    number = "7",
    pages = "071101",
    year = "2017"
}

@article{LHCb:2016inz,
    author = {{LHCb collaboration}},
    collaboration = "$\mathrm{LHCb}$",
    title = "{Search for massive long-lived particles decaying semileptonically in the LHCb detector}",
    eprint = "1612.00945",
    archivePrefix = "arXiv",
    primaryClass = "hep-ex",
    reportNumber = "CERN-EP-2016-283, LHCB-PAPER-2016-047",
    doi = "10.1140/epjc/s10052-017-4744-6",
    journal = "Eur. Phys. J. C",
    volume = "77",
    number = "4",
    pages = "224",
    year = "2017"
}

@article{LHCb:2016buh,
    author = {{LHCb collaboration}},
    collaboration = "$\mathrm{LHCb}$",
    title = "{Search for Higgs-like bosons decaying into long-lived exotic particles}",
    eprint = "1609.03124",
    archivePrefix = "arXiv",
    primaryClass = "hep-ex",
    reportNumber = "LHCB-PAPER-2016-014, CERN-EP-2016-188",
    doi = "10.1140/epjc/s10052-016-4489-7",
    journal = "Eur. Phys. J. C",
    volume = "76",
    number = "12",
    pages = "664",
    year = "2016"
}

@article{LHCb:2015ujr,
    author = {{LHCb collaboration}},
    collaboration = "$\mathrm{LHCb}$",
    title = "{Search for long-lived heavy charged particles using a ring imaging Cherenkov technique at LHCb}",
    eprint = "1506.09173",
    archivePrefix = "arXiv",
    primaryClass = "hep-ex",
    reportNumber = "CERN-PH-EP-2105-139, LHCB-PAPER-2015-002",
    doi = "10.1140/epjc/s10052-015-3809-7",
    journal = "Eur. Phys. J. C",
    volume = "75",
    number = "12",
    pages = "595",
    year = "2015"
}

@article{LHCb:2014jgs,
    author = {{LHCb collaboration}},
    collaboration = "$\mathrm{LHCb}$",
    title = "{Search for long-lived particles decaying to jet pairs}",
    eprint = "1412.3021",
    archivePrefix = "arXiv",
    primaryClass = "hep-ex",
    reportNumber = "LHCB-PAPER-2014-062, CERN-PH-EP-2014-291",
    doi = "10.1140/epjc/s10052-015-3344-6",
    journal = "Eur. Phys. J. C",
    volume = "75",
    number = "4",
    pages = "152",
    year = "2015"
}

@article{LHCb:2014osd,
    author = {{LHCb collaboration}},
    collaboration = "$\mathrm{LHCb}$",
    title = "{Search for Majorana neutrinos in $B^- \to \pi^+\mu^-\mu^-$ decays}",
    eprint = "1401.5361",
    archivePrefix = "arXiv",
    primaryClass = "hep-ex",
    reportNumber = "CERN-PH-EP-2014-002, LHCB-PAPER-2013-064",
    doi = "10.1103/PhysRevLett.112.131802",
    journal = "Phys. Rev. Lett.",
    volume = "112",
    number = "13",
    pages = "131802",
    year = "2014"
}

@article{LHCb:2008vvz,
    author = "Alves, Jr., A. Augusto and others",
    collaboration = "$\mathrm{LHCb}$",
    title = "{The LHCb Detector at the LHC}",
    reportNumber = "LHCb-DP-2008-001",
    doi = "10.1088/1748-0221/3/08/S08005",
    journal = "JINST",
    volume = "3",
    pages = "S08005",
    year = "2008"
}

@article{Knapen:2017kly,
    author = "Knapen, Simon and Lou, Hou Keong and Papucci, Michele and Setford, Jack",
    title = "{Tracking down Quirks at the Large Hadron Collider}",
    eprint = "1708.02243",
    archivePrefix = "arXiv",
    primaryClass = "hep-ph",
    doi = "10.1103/PhysRevD.96.115015",
    journal = "Phys. Rev. D",
    volume = "96",
    number = "11",
    pages = "115015",
    year = "2017"
}

@inproceedings{Batell:2022pzc,
    author = "Batell, Brian and Low, Matthew and Neil, Ethan T. and Verhaaren, Christopher B.",
    title = "{Review of Neutral Naturalness}",
    booktitle = "{Snowmass 2021}",
    eprint = "2203.05531",
    archivePrefix = "arXiv",
    primaryClass = "hep-ph",
    reportNumber = "PITT-PACC-2205",
    month = "3",
    year = "2022"
}

@article{Farina:2017cts,
    author = "Farina, Marco and Low, Matthew",
    title = "{Constraining Quirky Tracks with Conventional Searches}",
    eprint = "1703.00912",
    archivePrefix = "arXiv",
    primaryClass = "hep-ph",
    doi = "10.1103/PhysRevLett.119.111801",
    journal = "Phys. Rev. Lett.",
    volume = "119",
    number = "11",
    pages = "111801",
    year = "2017"
}

@article{Kang:2008ea,
    author = "Kang, Junhai and Luty, Markus A.",
    title = "{Macroscopic Strings and 'Quirks' at Colliders}",
    eprint = "0805.4642",
    archivePrefix = "arXiv",
    primaryClass = "hep-ph",
    doi = "10.1088/1126-6708/2009/11/065",
    journal = "JHEP",
    volume = "11",
    pages = "065",
    year = "2009"
}

@article{Kribs:2009fy,
    author = "Kribs, Graham D. and Roy, Tuhin S. and Terning, John and Zurek, Kathryn M.",
    title = "{Quirky Composite Dark Matter}",
    eprint = "0909.2034",
    archivePrefix = "arXiv",
    primaryClass = "hep-ph",
    reportNumber = "FERMILAB-PUB-09-425-T",
    doi = "10.1103/PhysRevD.81.095001",
    journal = "Phys. Rev. D",
    volume = "81",
    pages = "095001",
    year = "2010"
}

@article{Cai:2008au,
    author = "Cai, Haiying and Cheng, Hsin-Chia and Terning, John",
    title = "{A Quirky Little Higgs Model}",
    eprint = "0812.0843",
    archivePrefix = "arXiv",
    primaryClass = "hep-ph",
    doi = "10.1088/1126-6708/2009/05/045",
    journal = "JHEP",
    volume = "05",
    pages = "045",
    year = "2009"
}

@article{Burdman:2006tz,
    author = "Burdman, Gustavo and Chacko, Z. and Goh, Hock-Seng and Harnik, Roni",
    title = "{Folded supersymmetry and the LEP paradox}",
    eprint = "hep-ph/0609152",
    archivePrefix = "arXiv",
    reportNumber = "SLAC-PUB-12115",
    doi = "10.1088/1126-6708/2007/02/009",
    journal = "JHEP",
    volume = "02",
    pages = "009",
    year = "2007"
}

@article{Burdman:2008ek,
    author = "Burdman, Gustavo and Chacko, Z. and Goh, Hock-Seng and Harnik, Roni and Krenke, Christopher A.",
    title = "{The Quirky Collider Signals of Folded Supersymmetry}",
    eprint = "0805.4667",
    archivePrefix = "arXiv",
    primaryClass = "hep-ph",
    reportNumber = "UCB-PTH-08-09, SLAC-PUB-13251",
    doi = "10.1103/PhysRevD.78.075028",
    journal = "Phys. Rev. D",
    volume = "78",
    pages = "075028",
    year = "2008"
}

@article{Craig:2016kue,
    author = "Craig, Nathaniel and Knapen, Simon and Longhi, Pietro and Strassler, Matthew",
    title = "{The Vector-like Twin Higgs}",
    eprint = "1601.07181",
    archivePrefix = "arXiv",
    primaryClass = "hep-ph",
    doi = "10.1007/JHEP07(2016)002",
    journal = "JHEP",
    volume = "07",
    pages = "002",
    year = "2016"
}

@article{Craig:2015pha,
    author = "Craig, Nathaniel and Katz, Andrey and Strassler, Matt and Sundrum, Raman",
    title = "{Naturalness in the Dark at the LHC}",
    eprint = "1501.05310",
    archivePrefix = "arXiv",
    primaryClass = "hep-ph",
    reportNumber = "UMD-PP-014-028, CERN-PH-TH-2014-263",
    doi = "10.1007/JHEP07(2015)105",
    journal = "JHEP",
    volume = "07",
    pages = "105",
    year = "2015"
}

@article{Sha:2024hzq,
    author = "Sha, Qiyu and Murnane, Daniel and Fieg, Max and Tong, Shelley and Zakharyan, Mark and Fang, Yaquan and Whiteson, Daniel",
    title = "{Learning to Reconstruct Quirky Tracks}",
    eprint = "2410.00269",
    archivePrefix = "arXiv",
    primaryClass = "hep-ex",
    month = "9",
    year = "2024"
}

@article{Evans:2018jmd,
    author = "Evans, Jared A. and Luty, Markus A.",
    title = "{Stopping Quirks at the LHC}",
    eprint = "1811.08903",
    archivePrefix = "arXiv",
    primaryClass = "hep-ph",
    doi = "10.1007/JHEP06(2019)090",
    journal = "JHEP",
    volume = "06",
    pages = "090",
    year = "2019"
}

@article{Feng:2024zgp,
    author = "Feng, Jonathan L. and Li, Jinmian and Liao, Xufei and Ni, Jian and Pei, Junle",
    title = "{Discovering quirks through timing at FASER and future forward experiments at the LHC}",
    eprint = "2404.13814",
    archivePrefix = "arXiv",
    primaryClass = "hep-ph",
    doi = "10.1007/JHEP06(2024)197",
    journal = "JHEP",
    volume = "06",
    pages = "197",
    year = "2024"
}

@article{CMS:2017kku,
    author = "Sirunyan, A. M. and others",
    collaboration = "CMS",
    title = "{Search for decays of stopped exotic long-lived particles produced in proton-proton collisions at $\sqrt{s}=$ 13 TeV}",
    eprint = "1801.00359",
    archivePrefix = "arXiv",
    primaryClass = "hep-ex",
    reportNumber = "CMS-EXO-16-004, CERN-EP-2017-330",
    doi = "10.1007/JHEP05(2018)127",
    journal = "JHEP",
    volume = "05",
    pages = "127",
    year = "2018"
}

@article{CMS:2016ybj,
    collaboration = "CMS",
    title = "{Search for heavy stable charged particles with $12.9~\mathrm{fb}^{-1}$ of 2016 data}",
    reportNumber = "CMS-PAS-EXO-16-036",
    year = "2016"
}

@article{ATLAS:2016bek,
    author = "Aaboud, Morad and others",
    collaboration = "ATLAS",
    title = "{Search for new phenomena in final states with an energetic jet and large missing transverse momentum in $pp$ collisions at $\sqrt{s}=13$  TeV using the ATLAS detector}",
    eprint = "1604.07773",
    archivePrefix = "arXiv",
    primaryClass = "hep-ex",
    reportNumber = "CERN-EP-2016-075",
    doi = "10.1103/PhysRevD.94.032005",
    journal = "Phys. Rev. D",
    volume = "94",
    number = "3",
    pages = "032005",
    year = "2016"
}

@article{Akiba:2024now,
    author = "Akiba, K. and others",
    title = "{The LHCb VELO Upgrade module construction}",
    eprint = "2404.13615",
    archivePrefix = "arXiv",
    primaryClass = "physics.ins-det",
    reportNumber = "LHCb-DP-2024-001, LHCb-PAPER-2024-001",
    doi = "10.1088/1748-0221/19/06/P06023",
    journal = "JINST",
    volume = "19",
    number = "06",
    pages = "P06023",
    year = "2024"
}

@article{Buchanan:2022iib,
    author = "Buchanan, E. and others",
    title = "{Spatial resolution and efficiency of prototype sensors for the LHCb VELO Upgrade}",
    eprint = "2201.12130",
    archivePrefix = "arXiv",
    primaryClass = "physics.ins-det",
    doi = "10.1088/1748-0221/17/06/P06038",
    journal = "JINST",
    volume = "17",
    number = "06",
    pages = "P06038",
    year = "2022"
}

@article{Li:2020aoq,
    author = "Li, Jinmian and Li, Tianjun and Pei, Junle and Zhang, Wenxing",
    title = "{The quirk trajectory}",
    eprint = "2002.07503",
    archivePrefix = "arXiv",
    primaryClass = "hep-ph",
    doi = "10.1140/epjc/s10052-020-8209-y",
    journal = "Eur. Phys. J. C",
    volume = "80",
    number = "7",
    pages = "651",
    year = "2020"
}

@article{deAguiarFrancisco:2024bzg,
    author = "de Aguiar Francisco, Oscar Augusto",
    collaboration = "$\mathrm{LHCb~VELO~Upgrade~II~group}$",
    title = "{The LHCb VELO Upgrade II: Design and development of the readout electronics}",
    doi = "10.1016/j.nima.2024.169238",
    journal = "Nucl. Instrum. Meth. A",
    volume = "1063",
    pages = "169238",
    year = "2024"
}

@article{Okun:1979tgr,
    author = "Okun, L. B.",
    title = "{THETONS}",
    reportNumber = "DESY-L-TRANS-243",
    journal = "Pisma Zh. Eksp. Teor. Fiz.",
    volume = "31",
    pages = "156--159",
    year = "1979"
}

@article{Okun:1980mu,
    author = "Okun, L. B.",
    title = "{THETA PARTICLES}",
    reportNumber = "ITEP-6-1980",
    doi = "10.1016/0550-3213(80)90439-3",
    journal = "Nucl. Phys. B",
    volume = "173",
    pages = "1--12",
    year = "1980"
}

@article{LHCb:2018coe,
    author = "Bediaga, I. and others",
    collaboration = "$\mathrm{LHCb}$",
    title = "{Physics case for an LHCb Upgrade II - Opportunities in flavour physics, and beyond, in the HL-LHC era}",
    eprint = "1808.08865",
    archivePrefix = "arXiv",
    primaryClass = "hep-ex",
    reportNumber = "CERN-LHCC-2018-027, LHCB-PUB-2018-009",
    month = "8",
    year = "2018"
}

@misc{LHCb-Outreach:2025run,
    author = "Pietrzyk, B. and others",
    collaboration = "$\mathrm{LHCb}$",
    title = "{End of the 2025 proton-proton collision run}",
    howpublished = "\url{https://lhcb-outreach.web.cern.ch/2025/11/11/end-of-the-2025-proton-proton-collision-run/}",
    year = "2025",
    month = "11",
    note = "LHCb collected 11.8 fb$^{-1}$ of pp data in 2025"
}

@techreport{Coco:2683878,
      author        = "Coco, Victor and Akiba, Kazu and Back, John and Bertella,
                       Claudia and Bitadze, Alexander and Boente García, Oscar
                       and Bogdanova, Galina and Borghi, Silvia and Bowcock,
                       Themis and Bridges, Kieran and Brock, Matthew and Buytaert,
                       Jan and Byczinski, Wiktor and Carroll, John and Collins,
                       Paula and Dall'Occo, Elena and de Capua, Stefano and de
                       Bruyn, Kristof and de Roo, Krista and Dettori, Francesco
                       and di Canto, Angelo and Doets, Martin and Dosil Suarez,
                       Alvaro and Dreimanis, Karlis and Dumps, Raphael and Dutta,
                       Deepanwita and Eklund, Lars and Elvin, Andrew and Evans,
                       Tim and Fernández Prieto, Antonio and Ferro-Luzzi, Massi
                       and Flores, Leyre and Franco Lima, Vinicius and Franscisco,
                       Oscar and Freestone, Julian and Funk, Wolfgang and Fuzipeg,
                       Claire and Gallas Torreira, Abraham and Garcia Pardiñas,
                       Julián and García Plana, Beatriz and Gersabeck, Marco and
                       Gershon, Tim and Helena Mendes, Larissa and Hennessy, Karol
                       and Hulsbergen, Wouter and Hutchcroft, David and Hynds,
                       Daniel and Jalocha, Pawel and Jans, Eddy and John, Malcolm
                       and John, Dimitri and Jurik, Nathan and Ketel, Tjeerd and
                       Kopciewicz, Pawel and Kos, Johan and Kostiuk, Igor and
                       Kraan, Marco and Kuilman, Willem and Latham, Tom and
                       Leflat, Alexander and Lemos Cid, Edgar and Majewski, Maciej
                       and Marinho, Franciole and McCormick, Kevin and Merk,
                       Marcel and Meyer Garcia, Lucas and Miller, Graham and
                       Morris, Andrew and Munneke, Berend and Murray, Dónal and
                       Naik, Sneha and Nasteva, Irina and Oblakowska-Mucha,
                       Agnieszka and Otalora, Juan and Pérez Trigo, Eliseo and
                       Parkes, Chris and Pazos Álvarez, Antonio and Perry,
                       Michael and Rachwal, Bartlomej and Rinnert, Kurt and
                       Rodrigues, Gabriel and Roeland, Erno and Romero Vidal,
                       Antonio and Rovekamp, Joop and Sanchez Graz, Cristina and
                       Sanders, Freek and Scantlebury-Smead, Luke and Schiller,
                       Manuel and Schindler, Heinrich and Seman Bobulska, Dana and
                       Shears, Tara and Smith, Tony and Snoch, Aleksandra and
                       Švihra, Peter and Szumlak, Tomasz and Vázquez Regueiro,
                       Pablo and Van Beuzekom, Martin and van Overbeek, Martijn
                       and van Dongen, Jesse and Vazquez Sierra, Carlos and
                       Velthuis, Jaap and Vieites Diaz, Maria and Volkov, Vladimir
                       and Whitley, Mark and Williams, Mark",
      title         = "{Velo Upgrade Module Nomenclature}",
      institution   = "CERN",
      reportNumber  = "LHCb-PUB-2019-008, CERN-LHCb-PUB-2019-008",
      address       = "Geneva",
      year          = "2019",
      url           = "https://cds.cern.ch/record/2683878",
}

@article{Adinolfi:2012qfa,
    author = "Adinolfi, M. and others",
    title = "{Performance of the LHCb RICH detector at the LHC}",
    journal = "Eur. Phys. J. C",
    volume = "73",
    pages = "2431",
    year = "2013",
    eprint = "1211.6759",
    archivePrefix = "arXiv",
    primaryClass = "physics.ins-det",
    doi = "10.1140/epjc/s10052-013-2431-9"
}

@article{Feldman:1997qc,
    author = "Feldman, Gary J. and Cousins, Robert D.",
    title = "{A Unified approach to the classical statistical analysis of small signals}",
    eprint = "physics/9711021",
    archivePrefix = "arXiv",
    doi = "10.1103/PhysRevD.57.3873",
    journal = "Phys. Rev. D",
    volume = "57",
    pages = "3873--3889",
    year = "1998"
}

@misc{AtlasLumiRun3,
  author       = {{ATLAS Collaboration}},
  title        = {{ATLAS Luminosity Public Results Run 3}},
  howpublished = {\url{https://twiki.cern.ch/twiki/bin/view/AtlasPublic/LuminosityPublicResultsRun3}},
  year         = {2025},
  note         = {Accessed: 2025-12-07}
}

@misc{CMSLumiPublic,
  author       = {{CMS Collaboration}},
  title        = {{CMS Luminosity Public Results}},
  howpublished = {\url{https://twiki.cern.ch/twiki/bin/view/CMSPublic/LumiPublicResults}},
  year         = {2025},
  note         = {Accessed: 2025-12-07}
}

@article{Curtin:2025ksm,
    author = "Curtin, David and Dreyer, Sascha and Costa, Max Fust{\'e} and Heim, Sarah and Kasieczka, Gregor and Moureaux, Louis and Rousso, David and Shih, David and Sommerhalder, Manuel",
    title = "{Quirk SUEP}",
    eprint = "2506.11192",
    archivePrefix = "arXiv",
    primaryClass = "hep-ph",
    reportNumber = "DESY-25-081-0",
    month = "6",
    year = "2025"
}

@article{Condren:2025czc,
    author = "Condren, Levi and Whiteson, Daniel",
    title = "{Finding Unexpected Non-Helical Tracks}",
    eprint = "2509.08878",
    archivePrefix = "arXiv",
    primaryClass = "hep-ex",
    month = "9",
    year = "2025"
}

@article{Forsyth:2025wks,
    author = "Forsyth, Joshua and Low, Matthew and Tenney, Carson and Verhaaren, Christopher B.",
    title = "{Visible collider signals of natural quirks}",
    eprint = "2504.02940",
    archivePrefix = "arXiv",
    primaryClass = "hep-ph",
    doi = "10.1103/34xj-vz82",
    journal = "Phys. Rev. D",
    volume = "111",
    number = "11",
    pages = "115021",
    year = "2025"
}

@article{Li:2021tsy,
    author = "Li, Jinmian and Pei, Junle and Ran, Longjie and Zhang, Wenxing",
    title = "{The quirk signal at FASER and FASER 2}",
    eprint = "2108.06748",
    archivePrefix = "arXiv",
    primaryClass = "hep-ph",
    doi = "10.1007/JHEP12(2021)109",
    journal = "JHEP",
    volume = "12",
    pages = "109",
    year = "2021"
}

@article{Li:2019wce,
    author = "Li, Jinmian and Li, Tianjun and Pei, Junle and Zhang, Wenxing",
    title = "{Uncovering quirk signal via energy loss inside tracker}",
    eprint = "1911.02223",
    archivePrefix = "arXiv",
    primaryClass = "hep-ph",
    doi = "10.1103/PhysRevD.102.056006",
    journal = "Phys. Rev. D",
    volume = "102",
    number = "5",
    pages = "056006",
    year = "2020"
}
